\documentclass{aa}

\usepackage{graphicx}
\usepackage{txfonts}
\usepackage{natbib}
\bibpunct{(}{)}{;}{a}{}{,}

\begin{document}

   \title{Estimating turbulent velocities in the elliptical galaxies \\ NGC 5044 and NGC 5813}


   \author{J. de Plaa\inst{1} \and
           I. Zhuravleva\inst{2} \and
	   N. Werner\inst{3} \and
	   J. S. Kaastra\inst{1,4} \and
	   E. Churazov\inst{2,5} \and
	   R. K. Smith\inst{6} \and \\
	   A. J. J. Raassen\inst{1} \and
	   Y. G. Grange\inst{1} 
	  }

   \institute{SRON Netherlands Institute for Space Research,
              Sorbonnelaan 2, 3584 CA Utrecht, The Netherlands\\
              \email{j.de.plaa@sron.nl}
         \and 
	      Max-Planck-Institut f\"ur Astrophysik, 
	      Karl-Schwarzschild-Strasse 1, 85741 Garching, Germany     
	 \and
	      Kavli Institute for Particle Astrophysics and Cosmology,
	      Stanford University, 382 via Pueblo Mall, Stanford, CA 94305-4060, USA
	 \and 
              Astronomical Institute, Utrecht University,
	      PO Box 80000, 3508 TA Utrecht, The Netherlands
	 \and 
	      Space Research Institute (IKI), 
	      Profsoyuznaya 84/32, Moscow 117810, Russia
	 \and
	      Harvard-Smithonian Center for Astrophysics, 
	      60 Garden St., Cambridge, MA 02138, USA     
             }

   \date{Received 4 November 2011 / Accepted 22 December 2011}

   \abstract
   {The interstellar and intra-cluster medium in giant elliptical galaxies and clusters of galaxies is often assumed 
   to be in hydrostatic equilibrium. Numerical simulations, however, show that about 5--30\% of the pressure in a cluster    
   is provided by turbulence induced by, for example, the central active galactic nucleus and merger activity.}
   {We aim to put constraints on the turbulent velocities and the turbulent pressure in the intra-cluster medium of 
   the giant elliptical galaxies \object{NGC 5044} and \object{NGC 5813} using \textit{XMM-Newton} Reflection Grating 
   Spectrometer (RGS) observations.}
   {The magnitude of the turbulence is estimated using the \ion{Fe}{XVII} lines at 15.01 \AA, 17.05 \AA, and 17.10 \AA~in
   the RGS spectra. At low turbulent velocities, the gas becomes optically thick in the 15.01 \AA~line due to resonant
   scattering, while the 17 \AA~lines remain optically thin. By comparing the $(I_{17.05} + I_{17.10})/I_{15.01}$ line
   ratio from RGS with simulated line ratios for different Mach numbers, the level of turbulence is constrained. The
   measurement is, however, limited by the systematic uncertainty in the line ratio for an optically thin plasma, 
   which is about 20--30\%.}
   {We find that the $(I_{17.05} + I_{17.10})/I_{15.01}$ line ratio in \object{NGC 5813} is significantly higher 
   than in \object{NGC 5044}. This difference can be explained by a higher level of turbulence in NGC 5044. The best
   estimates for the turbulent velocities using resonant scattering and upper limits from the line widths, are
   320 $< V_{\mathrm{turb}} <$ 720 km s$^{-1}$ for NGC 5044 and 
   140 $< V_{\mathrm{turb}} <$ 540 km s$^{-1}$ for NGC 5813 at the 90\% confidence limit.}
   {The high turbulent velocities and the fraction of the turbulent pressure support of $>$40\% in NGC 5044,
   assuming isotropic turbulence, confirm that it is a highly disturbed system, probably due to an off-axis merger. The turbulent pressure 
   support in NGC 5813 is more modest at 15--45\%. The $(I_{17.05} + I_{17.10})/I_{15.01}$ line ratio in an optically 
   thin plasma, calculated using AtomDB v2.0.1, is 2$\sigma$ above the ratio measured in NGC 5044, which cannot be 
   explained by resonant scattering. This shows that the discrepancies between theoretical, laboratory, and astrophysical
   data on \ion{Fe}{XVII} lines need to be reduced to improve the accuracy of the determination of turbulent velocities 
   using resonant scattering.}

   
   \maketitle

\section{Introduction}

The hot X-ray emitting plasma in giant elliptical galaxies and clusters of galaxies is often assumed to be in hydrostatic equilibrium. In addition to the thermal pressure, there are, however, other processes which could contribute significantly to the total pressure of the plasma. One of these sources is turbulence induced by merger and AGN activity \citep[e.g.][]{loeb1994,roettiger1996,boehringer1993}. Recent estimates from numerical simulations suggest that bubbles blown by AGN can drive turbulence up to velocities of about 500 km s$^{-1}$ \citep[e.g.][]{brueggen2005,heinz2010}. In cluster mergers, the turbulence can go up to about 1000 km s$^{-1}$ and higher \citep[e.g.][]{dolag2005,lau2009}. Typically, turbulence is thought to provide about 5--30\% of the pressure support in clusters.   

Current X-ray observatories do not have the spectral resolution to measure turbulent velocities in the ICM directly by measuring the width of the emission lines. The Reflection Grating Spectrometers \citep[RGS,][]{herder2001} aboard {\it XMM-Newton} have the necessary spectral resolution, but because the spectrometers are slitless, the observed lines are blurred due to the extended nature of the source. Despite this complication, attempts have been made to put constraints on the turbulent velocities in groups and clusters of galaxies from the width of the lines in RGS spectra \citep{sanders2011}. Upper limits ranging from 200 km s$^{-1}$ up to 1500 km s$^{-1}$ were found in a sample of 62 clusters, groups and elliptical galaxies. 

Apart from velocity broadening of lines, there are also indirect methods to constrain turbulent motions in the hot gas, like resonant scattering. Although the hot plasma is in general optically thin, it can become optically thick at the energies of strong resonance lines \citep{gilfanov1987}. Since the transition probability of such a line is large, it has a high cross section to absorb and subsequently re-emit X-ray photons that have the same energy as the line. Along the line of sight toward the core of an elliptical galaxy, resonance-line photons that are emitted in the dense central region have a high probability to be scattered out of the beam, because the optical depth is large. Due to the conservative nature of resonant scattering, the scattered photons are re-distributed to larger projected distances. This effect causes an increase of the surface brightness at the edge of the galaxy \citep[see also the review by][]{churazov2010}. The optical depth depends on the turbulent broadening of resonant lines, especially for heavy elements like iron. A comparison of the fluxes of optically thin and optically thick lines could be used to place constraints on the level of turbulence in galaxies. 

In principle, there are several suitable lines to study resonant scattering in the X-ray band \citep[see][for a list of these lines emitted by cool and hot clusters]{churazov2010}. \citet{churazov2004} use the Fe-K complex of lines at 6.7 keV to find a lower limit to the turbulence in the Perseus cluster. Independently, using the same data, \citet{gastaldello2004} reached similar conclusions. The spectral resolution of the current instruments at these energies is however low, which limits the accuracy of the measurement. At the moment, the most useful lines to study resonant scattering are the 15.01~\AA, 17.05~\AA~and 17.10~\AA~lines of \ion{Fe}{XVII} in the Fe-L complex, which are detectable in relatively cool elliptical galaxies with RGS \citep{xu2002,werner2009}. The 15.01~\AA~line has the largest oscillator strength ($f=2.31-2.73$, depending upon the calculation) of the lines and is most sensitive to resonant scattering. The oscillator strengths of the 17.05~\AA~and 17.10~\AA~are much lower, $f=0.12$ and $f=10^{-8}$ respectively. Since all three lines are emitted by \ion{Fe}{XVII} ions, the ratio between the 15~\AA~and 17~\AA~lines is a good indicator for resonant scattering. In a sample of five elliptical galaxies, \citet{werner2009} find that the 15.01~\AA~line is suppressed in four out of five systems. Since the significance of the individual detections is low, 2--4$\sigma$ per galaxy, they could only estimate an upper limit for the turbulent velocity in \object{NGC 4636} which is 0.25 times the sound speed at the $\sim$90\% confidence level. In these studies, isotropic turbulence is assumed. However, these measurements are sensitive to the radial direction of the turbulence. If the direction of the motions are predominantly radial or tangential, it would change the results \citep{zhuravleva2011}.

Recently, long RGS exposures were obtained for the cool ($kT <$1 keV) elliptical galaxies \object{NGC 5044} and \object{NGC 5813}. These two galaxies have very different morphologies. NGC 5044 has bubbles which rise buoyantly, but are moved around in all directions due to sloshing motions induced by a recent merger \citep{david2009,gastaldello2009}. NGC 5813, on the other hand, has bubbles which rise nearly linearly along the NE-SW axis and does not show evidence of sloshing or recent merger interaction \citep{randall2011,tran2001,emsellem2007}. In the sample of \citet{werner2009}, NGC 5813 shows a hint of a presence of resonant scattering. The $(I_{17.05} + I_{17.10})/I_{15.01}$ line ratio is 1.99$\pm$0.34 in a 30 ks observation with respect to an expected ratio of 1.3, which suggests that the turbulent velocity is indeed low as is expected for a mildly disturbed system. We aim to further constrain the $(I_{17.05} + I_{17.10})/I_{15.01}$ line ratio for both galaxies using the deep RGS spectra and obtain better limits on the turbulent velocity. This would allow us to study the connection between the morphology of the ICM and the turbulent velocity in two very different environments.  

In this paper, we use a distance of 31.2 Mpc for NGC 5044 and 32.2 Mpc for NGC 5813 \citep{tonry2001}. At these distances, 1$^{\prime}$ corresponds to 9.1 kpc for NGC 5044 and 9.4 kpc for NGC 5813. Metal abundances are given with respect to proto-solar abundances by \citet{lodders2003} and uncertainties are given at the 68\% confidence level, unless stated otherwise.

\section{Data analysis}

In our analysis, we use all available RGS data on NGC 5044 and NGC 5813 that have exposures larger than 10 ks. Flares were discarded by filtering a light curve extracted from the outer regions of CCD 9. Bins with count rates above 0.25 counts/s were discarded. The observations were only modestly affected by soft-proton flares. For each observation, less than 20\% of the data was discarded due to flares. A summary of the observations used with their effective exposure time after flare filtering is provided in Table~\ref{tab:rgsexp}.

\subsection{RGS analysis}
\label{sec:rgs_analysis}

\begin{table}[t]
\caption{Effective exposure times for the XMM-Newton RGS observations after filtering.}
\label{tab:rgsexp}
\begin{center}
\begin{tabular}{lr|lr}
\multicolumn{2}{c}{\bf NGC 5044} 	& \multicolumn{2}{c}{\bf NGC 5813} \\
\hline\hline
{\bf ObsID}	& {\bf Exposure}& {\bf ObsID}	& {\bf Exposure} \\
\hline
0037950101	& 20.2 ks	& 0302460101    & 29.9 ks \\
0554680101	& 105.3 ks	& 0554680201	& 54.9 ks \\
		&		& 0554680301	& 52.6 ks \\
\hline
Total		& 125.5 ks	& Total		& 137.4 ks \\	
\hline	
\end{tabular}
\end{center}
\end{table}

RGS spectra of NGC 5044 and NGC 5813 were extracted using the SAS {\it XMM-Newton} analysis software version 10.0 \citep[see][for a detailed description of the analysis]{tamura2001a,deplaa2006}. We use the RGS model backgrounds provided by the SAS software as a best estimate for the background spectrum and we subtract it from the extracted source spectra. Since the suppression of the 15~\AA~line due to resonant scattering would be largest in the very centre of the group, we extract the spectrum from a 1$^{\prime}$ wide region (from $-0.5^{\prime}$ to $0.5^{\prime}$ with respect to the centre of the group) in the cross-dispersion direction of the RGS instrument.    

Since RGS is a slitless spectrometer, we need to take the line broadening due to the spatial extent of the source into account. The shape of the observed line profile depends on the surface brightness distribution of the source in the dispersion direction of RGS. In order to model the line broadening, a spatial profile is extracted from MOS1 data in a region corresponding to the selected RGS field of view. We put the profile on a wavelength grid using the relation $\Delta\lambda = 0.138~\AA~\Delta\theta$, where $\Delta\theta$ is the offset angle in arcminutes with respect to the centre of the source. To apply the effect of spatial broadening to the spectral model that is calculated during spectral fitting, we use this profile as a convolution kernel. The model spectrum is first convolved using the spatial profile, before we convolve it with the response matrix.
In the fit, we are able to optimise the width and centroid of the observed line profile by changing two parameters of the convolution kernel. The multiplication factor $s$ is the width of the resulting line profile with respect to the original profile from MOS1. The line centroid is represented by $\Delta\lambda^{\prime}$, which is the shift of the centroid in \AA ngstrom. The best-fit results of these parameters will be shown in addition to the other spectral parameters in the results section of this paper.

\subsection{Spectral modelling}
\label{sec:specmodel}

For the spectral fitting, we use the SPEX spectral fitting package \citep{kaastra1996}\footnote{See http://www.sron.nl/spex for more information}. We fit the RGS spectra with a single-temperature model of plasma in Collisional Ionisation Equilibrium (CIE). Since NGC 5044 shows evidence for multi-temperature structure \citep[see e.g.][]{tamura2003,grange2011}, we also fit its spectrum with a Gaussian Differential Emission Measure model \citep[\textit{gdem},][]{deplaa2006}. To measure the three lines of \ion{Fe}{XVII} at 15~\AA~and 17~\AA~independently of the spectral code, these lines have to be removed from the model and monochromatic lines have to take their place. In SPEX, this can be achieved by ignoring the \ion{Fe}{XVII} ion during the calculation of the line strengths. In Table~\ref{tab:lines}, we list all strong \ion{Fe}{XVII} lines in the $\sim$15~\AA~to 17~\AA~range. These lines are modelled with delta functions during fitting and also broadened by the spatial profile derived from MOS1. Delta lines can be used, because the natural width and the Doppler broadening of the lines are much smaller than the broadening due to the spatial extent of the source. We sum the lines at 17.05~\AA~and 17.10~\AA, because they are not resolved by RGS. 

In the 15--17 \AA~range, there are, of course, also weaker \ion{Fe}{XVII} lines that contribute to the flux, but which are not modelled if the \ion{Fe}{XVII} ion is ignored. Their contribution turns out to be limited. The sixth brightest line in the 15--17 \AA~range, which has a wavelength of 15.46 \AA~and is not listed in Table~\ref{tab:lines}, has an intensity which is at peak emission only 3.5\% of the intensity of the line at 15.01 \AA. Because of the small contribution of these weak lines and the contribution of stronger \ion{Fe}{XVII} lines outside the 15--17 \AA~range, we fix the temperature and Fe abundance to the best-fit value that we obtain from a fit where the \ion{Fe}{XVII} ion is included. 

\begin{table}
\caption{Wavelengths and oscillator strengths for all strong \ion{Fe}{XVII} lines in the 15 \AA~ to 17 \AA~ range.}
\label{tab:lines}
\begin{tabular}{@{}l@{~ }lrrrr@{}}
		& {\bf Transition}	& {\bf $\lambda$ (\AA)\tablefootmark{a}} & $f$\tablefootmark{b}	
		& $f$\tablefootmark{c}  & $f$\tablefootmark{d}	\\
\hline\hline
3C & 2s$^2$ 2p$^6$ $^1$S$_0$ -- 2s$^2$ 2p$^5$ 3d $^1$P$_1$ & 15.015 & 2.31	& 2.73		& 2.49\\
3D & 2s$^2$ 2p$^6$ $^1$S$_0$ -- 2s$^2$ 2p$^5$ 3d $^3$D$_1$ & 15.262 & 0.63	& 0.61		& 0.64\\
3F & 2s$^2$ 2p$^6$ $^1$S$_0$ -- 2s$^2$ 2p$^5$ 3s $^3$P$_1$ & 16.777 & 0.11	& 0.11		& 0.10\\
3G & 2s$^2$ 2p$^6$ $^1$S$_0$ -- 2s$^2$ 2p$^5$ 3s $^1$P$_1$ & 17.054 & 0.12	& 0.13		& 0.13\\ 
M2 & 2s$^2$ 2p$^6$ $^1$S$_0$ -- 2s$^2$ 2p$^5$ 3s $^3$P$_2$ & 17.097 &	& 5 10$^{-8}$	& 5 10$^{-8}$\\
\hline
\end{tabular}
\tablefoot{
\tablefoottext{a}{Wavelengths in \AA~according to NIST; http://physics.nist.gov/PhysRefData/ASD/lines\_form.html}
\tablefoottext{b}{Oscillator strengths as provided by \citet{shorer1979}; NIST http://physics.nist.gov/PhysRefData/ASD/lines\_form.html}
\tablefoottext{c}{Calculated oscillator strengths (D. Liedahl/SPEX)}
\tablefoottext{d}{AtomDB v2.0.1 \citep{loch2006}}
}
\end{table}

\begin{figure}[t]
\includegraphics[width=\columnwidth]{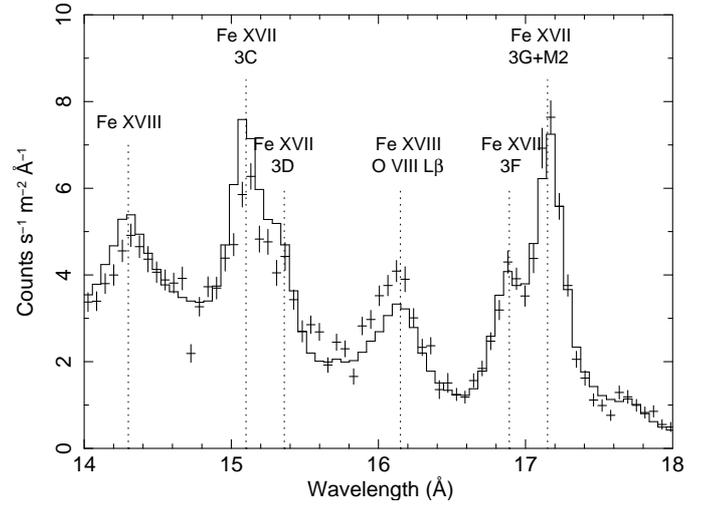}
\caption{RGS spectrum of NGC 5813 in the 14--18~\AA~band. The line properties of the labelled \ion{Fe}{XVII} transitions can be found in Table~\ref{tab:lines}.}
\label{fig:rgsspec}
\end{figure}

Fig.~\ref{fig:rgsspec} shows the fluxed RGS spectrum of NGC 5813 in the 14--18~\AA~band. The model fit does not take into account resonant scattering effects. The figure shows that the line at 15~\AA~(3C) is overestimated by the model. However, also the other lines from \ion{Fe}{XVIII} and \ion{O}{VIII} are not well fitted. These discrepancies are most likely due to uncertainties in the atomic data and the ionisation balance, which can lead to systematic errors up to 30\% in individual line strengths. Clearly, this is an issue for the \ion{Fe}{XVII} lines as well. Therefore, we discuss the uncertainties in the atomic data of \ion{Fe}{XVII} extensively in this paper.

\section{Modelling of resonant scattering}
\label{sec:resscat}

Modelling the resonant scattering effect that is observed in RGS spectra consists of two parts. First, the \ion{Fe}{XVII} line ratio $(I_{17.05} + I_{17.10})/I_{15.01}$ for an optically thin plasma should be determined. Since we cannot measure this ratio directly from our observations due to the yet unknown optical depth of the plasma in the 15 \AA~line, it is calculated using atomic data and spectral modelling. Secondly, we calculate how this ratio is affected by resonant scattering and turbulence in the ICM.

\subsection{The \ion{Fe}{XVII} line ratios in optically thin plasmas}
\label{sec:atomic}

\begin{figure}[t]
\includegraphics[width=\columnwidth]{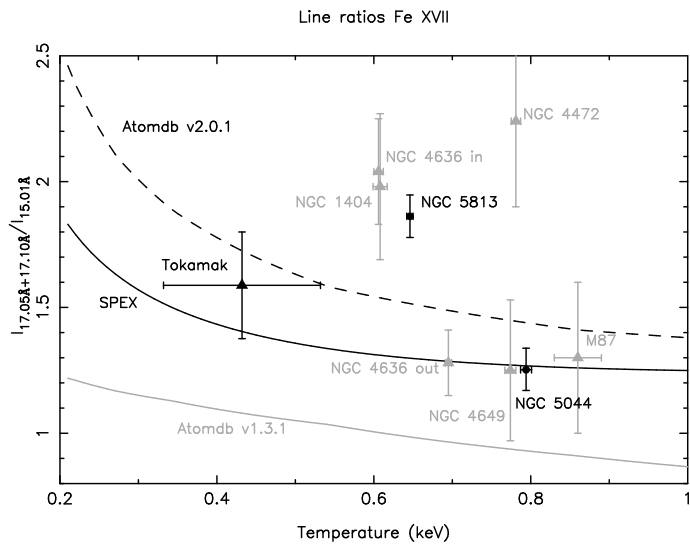}
\caption{Measured and theoretical values for the line ratio $(I_{17.05} + I_{17.10})/I_{15.01}$ as a function of temperature. The lines indicate the values predicted by the currently available spectral codes: AtomDB v1.3.1 ({\it grey}), SPEX ({\it black}) updated with \citet{doron2002}, and AtomDB v2.0.1 ({\it dashed}). The black triangle indicates the value found through laboratory measurements \citep{beiersdorfer2004}. The grey triangles show measurements of other ellipticals performed by \citet{werner2009} and \citet{werner2006b} (M87). For M87, the temperature determining the line ratio is assumed to be the coolest temperature in the two-temperature model. The black dots show our measurements for NGC 5044 and NGC 5813.}
\label{fig:ratio_all}
\end{figure}

The line intensities of the Ne-like \ion{Fe}{XVII} lines in the 15--18 \AA~range (see Table~\ref{tab:lines}) have been subject to a long standing debate in atomic physics \citep[part of the debate is summarised in, for example,][]{gu2007}. For many years, the two main groups performing laboratory experiments at the National Institute of Standards and Technology (NIST) and Laurence Livermore National Laboratory (LLNL) were unable to reach agreement on their measured \ion{Fe}{XVII} line ratios \citep[see, for example][]{beiersdorfer2003}. On the other side, theorists, using different methods and codes, also obtained ratios that were neither consistent with other theoretical calculations nor with the experimental data. Only recently, \citet{gillaspy2011} appear to have converged to experimental values that agree with the closest theoretical predictions within 10\%. However, given the history of this debate, this is likely not the final word on this issue.     

There are several publicly available spectral codes that attempt to model the line strengths of the \ion{Fe}{XVII} lines in astrophysical plasmas either based on theoretical calculations, experimental data, or both. In this work, we compare the results from AtomDB\footnote{Website: http://www.atomdb.org}, which is integrated in the widely used APEC model, with the SPEX code, which is based on MEKAL \citep{mewe95} and updated with the theoretical \ion{Fe}{XVII} calculations by \citet{doron2002}. The $(I_{17.05} + I_{17.10})/I_{15.01}$ ratios as a function of the temperature for the different codes are plotted in Fig.~\ref{fig:ratio_all}. As can be expected from the inconsistencies between the experimental atomic data and theoretical predictions, the spectral codes do not agree.

The codes based on older atomic data, AtomDB v1.3.1 and MEKAL (not shown in Fig.~\ref{fig:ratio_all}), both predict a line ratio of about 1.0 with a relatively small dependence on temperature, dropping from about 1.2 at 0.2 keV to 0.9 at 1.0 keV. However, comparisons between stellar flare observations performed by the {\it XMM-Newton} and {\it Chandra} grating spectrometers \citep[e.g.][]{mewe2001,ness2002}, laboratory measurements \citep[e.g.][]{beiersdorfer2004}, and theoretical calculations \citep[e.g.][]{doron2002} suggest that this ratio is probably underestimated. \citet{doron2002} found that di-electronic recombination (DR) and resonant excitation (RE) processes are dominant in the 0.4--0.6 keV temperature range, while these processes were only modelled using scaling laws in MEKAL. Because of this finding, SPEX was updated at that time to change the predicted \ion{Fe}{XVII} line strengths to the new values based on these theoretical calculations. The resulting $(I_{17.05} + I_{17.10})/I_{15.01}$ ratio is shown as a black line in Fig.~\ref{fig:ratio_all}. This ratio is considerably higher than the results from the MEKAL and AtomDB v1.3.1 codes. It drops from a ratio of about 1.9 at 0.2 keV to 1.25 at 1.0 keV. 

The results obtained by \citet{doron2002} are, however, not without problems. Although they were right to consider recombination and ionisation in their model, neither their distorted wave calculations nor the ionisation balance that they adopted \citep{mazzotta1998} are state-of-the-art for these lines or ions. For example, the single-ion model of \citet{doron2002} predicts a value for the $(I_{16.78} + I_{17.05} + I_{17.10})/I_{15.01}$ ratio that is always $<$2 for temperatures below 1 keV. But according to new experimental and theoretical data \citep{gillaspy2011}, this ratio should be $>$2 at these temperatures. The results from the 3-ion model (\ion{Fe}{XVI}, \ion{Fe}{XVII}, and \ion{Fe}{XVIII}) of \citet{doron2002} are more consistent with the experimental values, while they should not be, because the ionisation and recombination processes included in this model do not take place in the experimental setup. Electron beam ion trap (EBIT) experiments predominantly contain ions from the same kind, i.e. \ion{Fe}{XVII}. As \citet{doron2002} showed in the 3-ion model, the presence of the neighbouring ions (\ion{Fe}{XVI} and \ion{Fe}{XVIII}), as in astrophysical sources, increases the population of the levels of the 16.78~\AA, 17.05~\AA, and 17.10~\AA~lines, thus enhancing the $(I_{16.78} + I_{17.05} + I_{17.10})/I_{15.01}$ ratio with respect to laboratory measurements. Although this effect is genuine and expected in astrophysical plasmas, their predicted values using the single-ion model are not consistent with the laboratory values of \citet{gillaspy2011}.   

Recently, AtomDB v2.0.1 was released \citep{foster2010} containing major updates of the atomic data and new calculations of the line formation processes. The AtomDB developers chose to use more recent calculations \citep{badnell2006b,badnell2006a}\footnote{http://amdpp.phys.strath.ac.uk/tamoc/DATA/} based on the code by \citet{badnell1986}, which should be more consistent with the experimental data and also includes ionisation and recombination processes, like in \citet{doron2002}. As shown by the dashed line in Fig.~\ref{fig:ratio_all}, the predicted ratio from AtomDB v2.0.1 has increased with respect to the other codes. The ratio drops from about 2.5 at 0.2 keV to 1.4 at 1.0 keV.

Comparing the predicted ratios with laboratory experiments or astrophysical observations is not straightforward, because it is difficult to determine to what extent resonant scattering is acting in the experimental setup or in the astrophysical source. The tokamak measurements reported by \citet{beiersdorfer2004} are believed not to be affected by resonant scattering. The weighted average of their measured data points is shown in Fig.~\ref{fig:ratio_all}. The values are consistent within errors with the updated spectral codes, but only marginally consistent with MEKAL and AtomDB v1.3.1. We also compare these ratios with the values measured in a small sample of giant elliptical galaxies \citep{werner2009,werner2006b}, which are shown in grey in Fig.~\ref{fig:ratio_all}. If a data point is above the calculated curve, then it is an indication of resonant scattering, because the 15~\AA~line can be suppressed through photons effectively scattering out of the beam when pointing to the core of the galaxy. The only data point taken next to a galaxy core is "NGC 4636 out". In principle, only for this point the 15~\AA~line could be enhanced by photons effectively scattered into the beam and lowering the ratio, but since the spectrum in this region is dominated by the emission near the core of the galaxy where there is no enhancement yet, this effect is very small. Otherwise, finding a ratio below the calculated values is very unlikely.  

The astrophysical data from elliptical galaxies presented here appear to favour the ratio found using SPEX above the AtomDB v2.0.1 value if we only take the statistical uncertainties into account. This allows us to favour the numerical value obtained by SPEX, but it does not mean that the physical background of the \citet{doron2002} model is correct. The older MEKAL and AtomDB v1.3.1 underestimate the ratio considerably, which could be explained by the lower accuracy of the estimations used to model the line formation processes. Although the SPEX predictions appear to be more consistent with measurements, a systematic uncertainty in the line ratio of about 20--30\% remains when we consider the difference between SPEX and AtomDB v2.0.1 (Fig.~\ref{fig:ratio_all}) and the uncertainty in the oscillator strength of the 15.015~\AA~line (Table~\ref{tab:lines}). This is the same level of uncertainty as discussed in \citet{beiersdorfer2003}.

\subsection{Modelling resonant scattering in the ICM} 

\begin{table}[t]
\caption{{\it Chandra} observations of NGC 5044 and NGC 5813.}
\label{tab:chandra}
\begin{center}
\begin{tabular}{l|rr|rr}
		& \multicolumn{2}{c}{{\bf NGC 5044}} 	& \multicolumn{2}{c}{{\bf NGC 5813}} \\
\hline\hline
ObsID		& 798		& 9399		& 5907		& 9517	\\
Exposure time	& 21 ks		& 84 ks		& 49 ks		& 100 ks \\
\hline
Total exposure  & \multicolumn{2}{c}{105 ks}	& \multicolumn{2}{c}{149 ks} \\
\hline
\end{tabular}
\end{center}
\end{table}

\begin{figure*}
\begin{center}
\includegraphics[width=1.01\columnwidth]{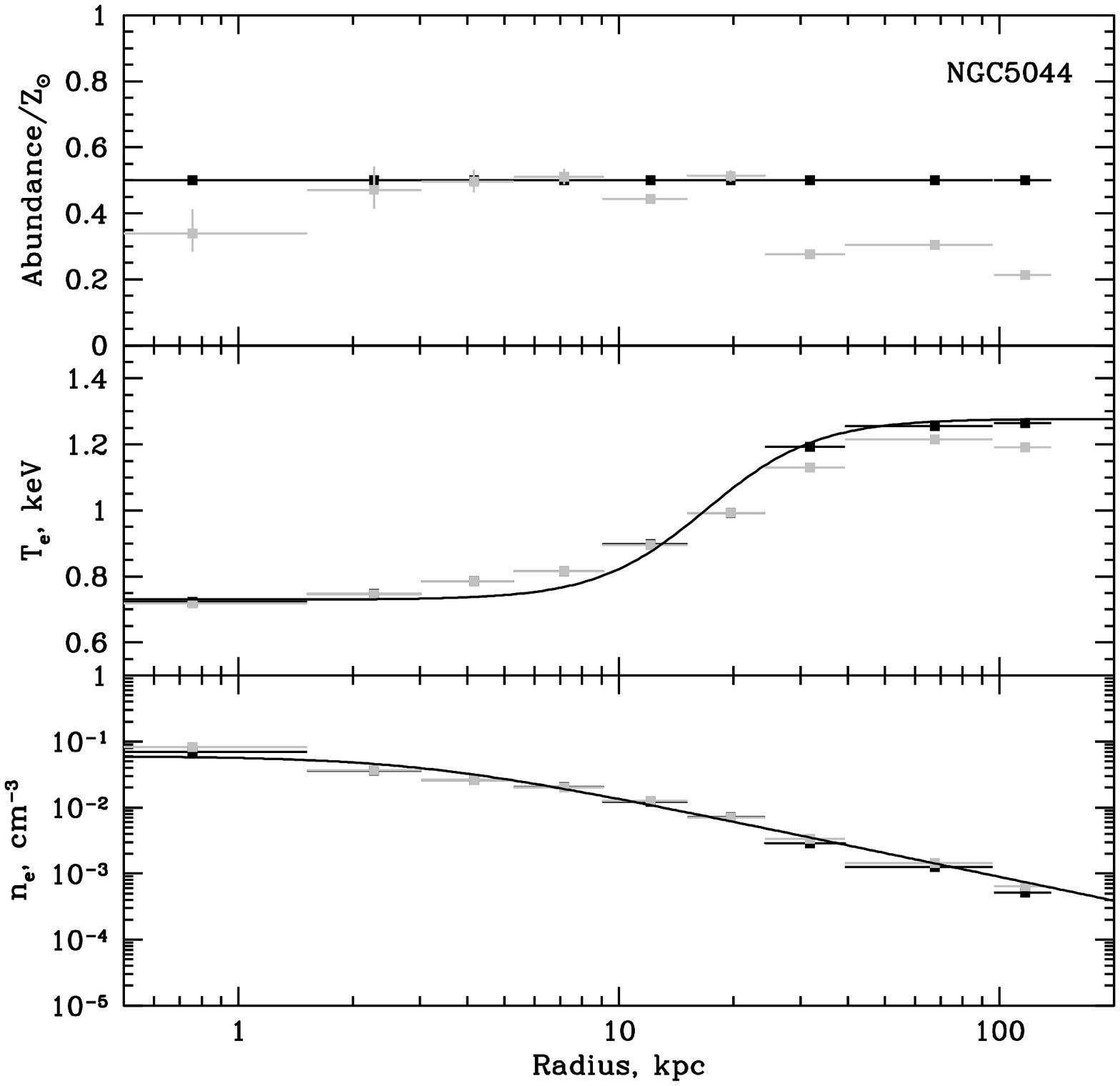}
\includegraphics[width=1.01\columnwidth]{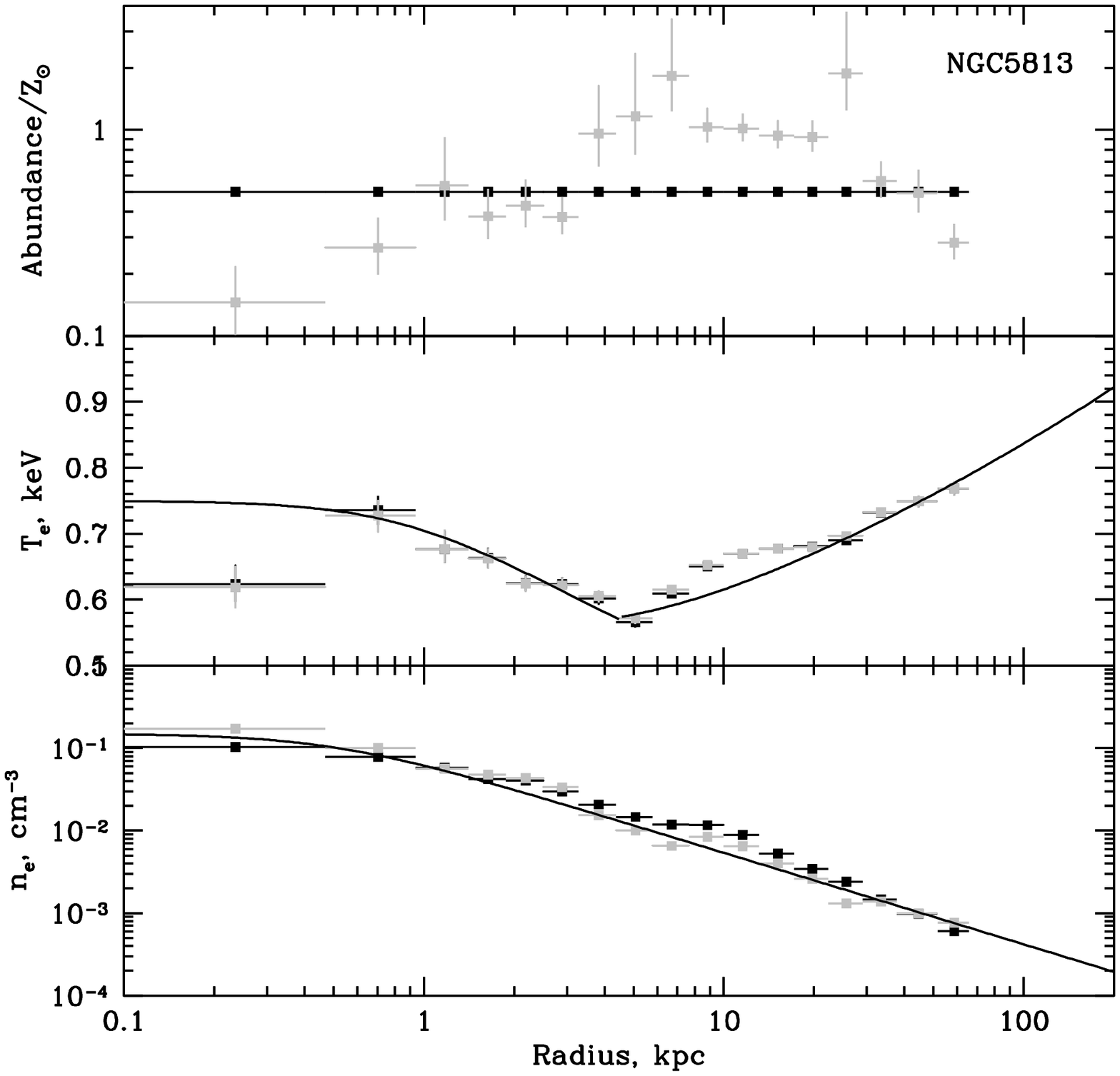}
\end{center}
\caption{Deprojected profiles of number electron density, temperature and Fe
  abundance in NGC5044 (left panel) and NGC5813 (right panel). Grey
  points: deprojected profiles with free abundance, black points:
  deprojected profiles with fixed abundance. Black curves: fits of
  density and temperature used in our calculations.}
\label{fig:profiles}
\end{figure*}

The modelling of resonant scattering in the 15~\AA~line in giant ellipticals is 
performed through Monte Carlo simulations. As an input to the simulations, we use 
deprojected density, temperature and Fe abundance profiles obtained with {\it Chandra}. 
The {\it Chandra} observations of NGC 5044 and NGC 5813 that we use, are listed in 
Table~\ref{tab:chandra}. The initial data processing is performed using the 
standard method described in \citet{vikhlinin2005} and using the latest 
calibration corrections. Subsequently, we deproject the observed spectrum 
to measure the temperature, density and abundance profiles of the plasma. 
The deprojection technique is described in detail
by \citet{churazov2003} and \citet{churazov2008}. In brief, we find the set
of 3D spectra in spherical shells that fit the observed spectra in concentric 
annuli best after projection. Assuming that the volume
emissivity in the outermost shell at all energies declines as a power
law with radius, we model the observed spectra as a linear combination
of spectra in spherical shells plus the contribution of the outermost
layers. The matrix that describes the projection of the shells into
annuli with the contribution of outer layers is inverted and the spectra in the
shells are calculated by applying the inverted matrix to the observed
spectra. The resulting spectra in every shell are fitted in XSPEC with
an APEC model \citep{smith2001}.

The radial profiles of the deprojected gas parameters in NGC 5044 and
NGC 5813 are shown in Fig.~\ref{fig:profiles}. One can see that
assuming a fixed abundance of 0.5 times Solar \citep{anders1989} does not
change the profiles of the density and temperature significantly. The black
curves show the fits of the density and temperature that we used as an input for
our Monte Carlo simulations. The number electron density in NGC 5044 is described as a
$\beta-$model with a normalisation of $n_0=0.06$ cm$^{-3}$, a core radius of $r_c=3$ kpc
and a $\beta=0.4$. The temperature profile (in keV) is parametrised as
\begin{equation}
T_{\mathrm{e}}(r) = 0.73\frac{1+1.75\left(\displaystyle \frac{r}{17}\right)^3}{1+\left(\displaystyle \frac{r}{17}\right)^3},
\end{equation}
where $r$ is in kpc. The number electron density in NGC 5813 is also described as a
$\beta-$model with a normalisation of $n_0=0.15$ cm$^{-3}$, a core radius of $r_c=0.5$ kpc
and a $\beta=0.37$. The temperature is described as
\begin{equation}
T_{\mathrm{e}}(r)= \left\{
\begin{array}{lc}
0.55\left[1+\left(\displaystyle\frac{r}{5}\right)^2\right]^{0.07} &  r \geq 5 ~\mathrm{kpc} \\
  0.75\left[1+\left(\displaystyle\frac{r}{1}\right)^2\right]^{-0.09} &  r < 5 ~\mathrm{kpc} \\
\end{array} \right.
\end{equation}

\subsubsection{Monte Carlo simulations}

In the Monte Carlo simulations, we calculate the optical depth $\tau$ in the centre of the line using
\begin{equation}
\tau=\int n_i \sigma_0 dl,
\end{equation}
where $l$ is the distance along the photon propagation direction,
$n_i$ is the number density of ions in the ground state of a given
transition and $\sigma_0$ is the cross section for scattering at the
centre of a resonant line. This cross-section is given by
\begin{equation}
\sigma_0=\frac{\sqrt{\pi}hr_{\mathrm{e}} cf}{\Delta E_{\mathrm{D}}},
\end{equation}
where $h$ is the Planck constant, $r_{\mathrm{e}}$ is the classical electron radius, $f$ the absorption
oscillator strength of a given atomic transition and $\Delta E_{\mathrm{D}}$ the
Doppler width of the line, which is defined as
\begin{equation}
\Delta E_{\mathrm{D}}=E_0\left(\frac{2kT_{\mathrm{e}}}{Am_{\mathrm{p}}c^2}+\frac{V^2_{\mathrm{turb}}}{c^2}\right)^{1/2}.
\end{equation}
Here, $E_0$ is the rest energy of the line, $T_{\mathrm{e}}$ is the electron
temperature, $A$ is the atomic mass of the ion, $m_p$ is the
proton mass and $c$ is the speed of light. Using the adiabatic sound
speed $c_s$, one can express the level of turbulence as a Mach number
$M=V_{\mathrm{turb}}/c_s$ \citep[see, e.g.,][]{churazov2004}, i.e.
\begin{equation}
\Delta E_{\mathrm{D}}=E_0\left[\frac{2kT_{\mathrm{e}}}{Am_{\mathrm{p}} c^2}(1+1.4AM^2)\right]^{1/2}.
\end{equation}
The calculated optical depths in the considered lines of both galaxies are shown in Table \ref{tab:optdepth} for
Mach numbers ranging from 0 to 0.75.

\begin{table}
\caption{Optical depths ($\tau$) in the \ion{Fe}{XVII} lines at 15.01~\AA~and 17.05~\AA~in the galaxies NGC 5044 and NGC 5813 calculated for Mach numbers 0, 0.25, 0.5 and 0.75.} 
\label{tab:optdepth}
\begin{center}
\begin{tabular}{l|cc|cc}
	& \multicolumn{2}{c}{\textbf{NGC 5044}} & \multicolumn{2}{c}{\textbf{NGC 5813}} \\
\hline\hline
Mach	&	15.01\AA	& 17.05\AA	& 15.01\AA	&	17.05\AA \\
\hline
0	&	3.8		& 0.2		& 6.3		&	0.4 \\
0.25	&	1.6		& 0.09		& 2.6		&	0.1  \\
0.50	&	0.8		& 0.05		& 1.4		&	0.08 \\
0.75	&	0.6		& 0.03		& 0.9		&	0.05 \\ 
\hline	
\end{tabular}
\end{center}
\end{table} 

\begin{figure*}[t]
\includegraphics[width=\columnwidth]{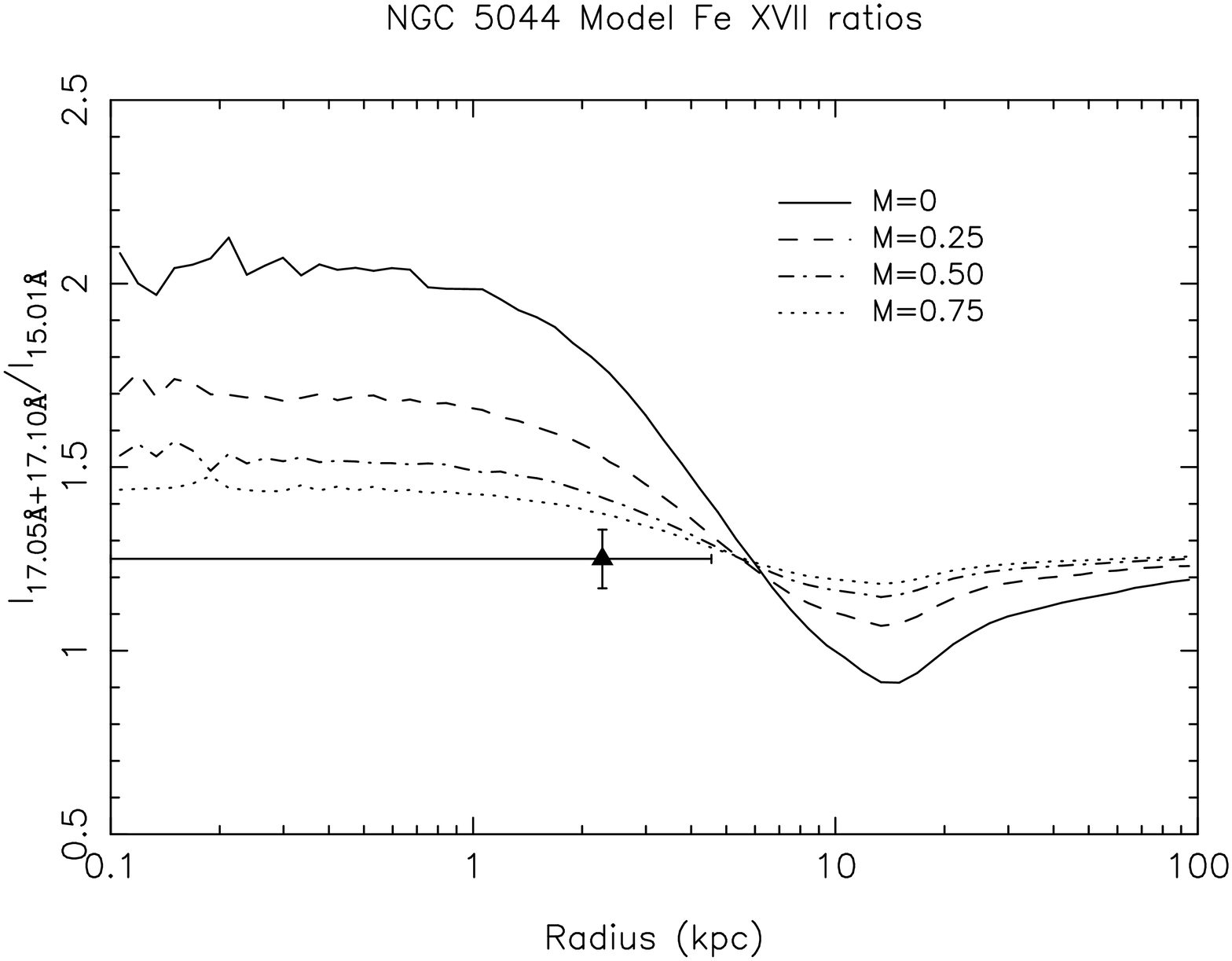}
\includegraphics[width=\columnwidth]{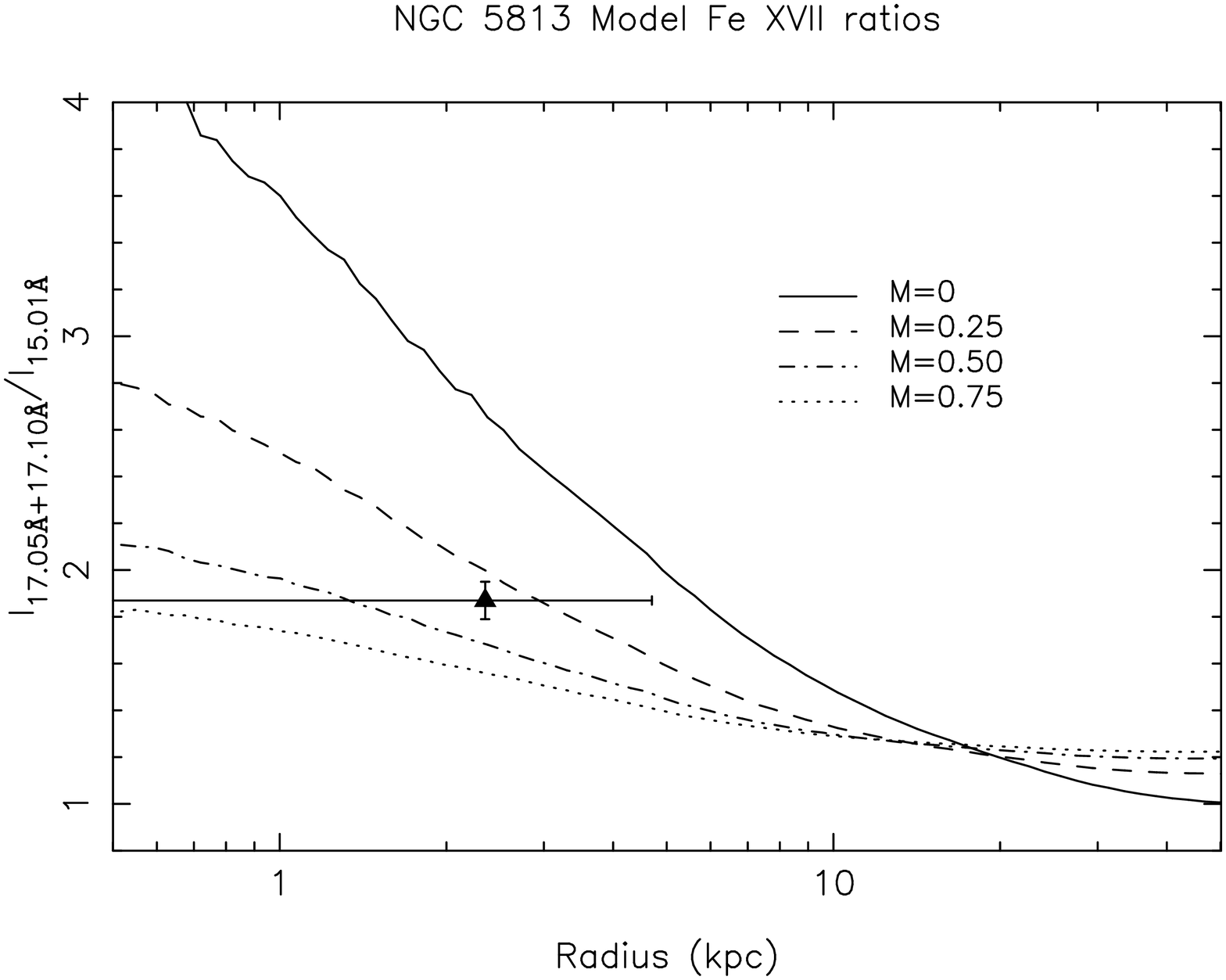}
\caption{\ion{Fe}{XVII} line ratios for NGC 5044 (\textit{left}) and NGC 5813 (\textit{right}). The lines give the predicted $(I_{17.05} + I_{17.10})/I_{15.01}$ line ratio for different Mach numbers (0, 0.25, 0.5, and 0.75) from Monte Carlo simulations assuming isotropic turbulence. Note that these profiles have been re-scaled to the optically-thin $(I_{17.05} + I_{17.10})/I_{15.01}$ line ratio from \citet{doron2002}. The data point in each plot is the measured line ratio derived from RGS.}
\label{fig:resscat} 
\end{figure*}

Assuming spherically symmetric models of galaxies, we then calculate the plasma line emissivities using 
APEC \citep{smith2001}. Line energies and oscillator strengths 
are taken from the AtomDB\footnote{http://www.atomdb.org/} \citep{loch2006} and NIST Databases\footnote{http://www.nist.gov/pml/data/asd.cfm}. The
ionisation balance (collisional equilibrium) is taken from \citet{bryans2009}.
Multiple resonant scatterings are calculated using a Monte Carlo approach,
which treats an individual act of resonant scattering in full
detail \citep[details of these simulations are described in e.g.][]{churazov2004,werner2009,zhuravleva2010,zhuravleva2011}.
In Fig.~\ref{fig:resscat}, we show the resulting ratio of the intensities in the \ion{Fe}{XVII} lines,
$(I_{17.05}+I_{17.10})/I_{15.01}$, for both elliptical galaxies and for Mach numbers between 0 and 0.75. 
One can see the effect of resonant scattering: in the core the line intensity is suppressed, while in the
surrounding regions the intensity of line increases. 

\section{Results}

\begin{table}[t]
\caption{RGS best-fit model results for NGC 5044 and NGC 5813 for an extraction region of 0.5$^{\prime}$ around the centre of the source. For NGC 5044 we also show the results from a multi-temperature {\it gdem} model. }
\label{tab:rgsresults}
\begin{center}
\begin{tabular}{lccc}
{\bf Parameters}		& {\bf NGC 5044}	& {\bf NGC 5044 {\it gdem}}	& {\bf NGC 5813}	\\
\hline\hline
$Y$ ($10^{70}$ m$^{-3}$)	& 8.9$\pm$0.3		& 8.8$\pm$0.3		& 1.79$\pm$0.12		\\
k$T$ (keV)			& 0.784$\pm$0.003	& 0.795$\pm$0.007	& 0.647$\pm$0.003	\\
$\sigma$			&			& 0.043$\pm$0.012	&			\\
O				& 0.40$\pm$0.03		& 0.42$\pm$0.03		& 0.49$\pm$0.05		\\
Ne				& 0.81$\pm$0.09		& 0.72$\pm$0.11		& 0.32$\pm$0.07		\\
Fe				& 0.56$\pm$0.02		& 0.59$\pm$0.03		& 0.70$\pm$0.05		\\
$s$				& 1.01$\pm$0.03		& 1.01$\pm$0.03		& 0.99$\pm$0.03		\\
$\Delta\lambda^{\prime}$ (m\AA)	& -1.1$\pm$3.1		& 0.2$\pm$3.2		& -1.2$\pm$2.7  	\\
\hline
C-stat / d.o.f.			& 959 / 518		& 955 / 517		& 1303 / 781	\\
\hline
\end{tabular}
\end{center}
\end{table}

We fit RGS 1 and RGS 2 first order spectra extracted from a region 0.5$^{\prime}$ around the centre of the group with a single temperature model. In addition, we also fit a multi-temperature {\it gdem} model to the spectrum of NGC 5044. Unfortunately, the spectral resolution deteriorates too quickly to derive meaningful results from the outer spatial regions in the cross-dispersion direction. The rotation angles of the RGS extraction regions with respect to the group emission is visualised in \citet{grange2011}. The best-fit parameters for a single-temperature model and the {\it gdem} model are listed in Table~\ref{tab:rgsresults}. Due to resonant scattering and the systematic uncertainties in the atomic data, the fits are not perfect, but acceptable for our purpose. The multi-temperature fit for NGC 5044 is comparable in quality to the single-temperature fit and yields parameter values that are consistent with the single-temperature results. The width of the lines due to the spatial extent of the source are well fitted by the spatial profile extracted from EPIC MOS, because the scale factor $s$ (see Section~\ref{sec:rgs_analysis}) is consistent with being 1 for both groups, which means that the width is the same as the width of the original profile. Also the wavelength, which depends on the position of the source, is estimated properly. The wavelength shift $\Delta\lambda^{\prime}$ is free in the fit and the best fit value is consistent with being 0. 

Another reason for the relatively high C-stat values that we obtain could be the presence of multi-temperature structure. The RGS data of NGC 5813, however, do not show significant evidence for that \citep{grange2011}. In NGC 5044, there are indications for multi-temperature structure, but the lower cut-off temperature is at 0.4 keV \citep{grange2011}, which means that there is no indication for temperatures below 0.4 keV. Fig.~\ref{fig:ratio_all} shows that the expected line ratios flatten out above temperatures of $\sim$0.5 keV, which just adds an uncertainty of $\lesssim$10\% to the optically-thin line ratio, which is small compared to the uncertainties in the atomic data. Since plasma below temperatures of $\sim$0.5 keV is extremely rare in cool-core clusters and giant ellipticals \citep{sanders2011b} and we do not see evidence for \ion{O}{VII} emission lines indicating temperatures below $\sim$0.4 keV, we will use the results from the single-temperature model.

To determine the $(I_{17.05} + I_{17.10})/I_{15.01}$ line ratio, we use the best-fit model shown in Table~\ref{tab:rgsresults} and ignore \ion{Fe}{XVII} in the spectral calculation. This removes the strongest \ion{Fe}{XVII} lines, which are listed in Table~\ref{tab:lines}, from the model. We therefore replace these five lines with four delta functions. Since we cannot resolve the 17.05~\AA~and 17.10~\AA~lines, they are modelled by a single delta function at the average wavelength of the two lines. Then, we fit the spectrum again with the temperature and Fe abundance fixed to their best-fit values, because these two parameters are likely to introduce bias due to the lack of \ion{Fe}{XVII} lines in the spectral model (See Sect.~\ref{sec:specmodel}). Other parameters, like the normalisation, the oxygen and neon abundances, the delta lines, and the spatial broadening parameters are allowed to optimise. From these fits, we obtain a $(I_{17.05} + I_{17.10})/I_{15.01}$ line ratio of 1.25$\pm$0.08 for NGC 5044 and 1.87$\pm$0.08 for NGC 5813. Also the ratio from the multi-temperature fit of NGC 5044 yields 1.25$\pm$0.08, which shows that the line ratio does not depend on the multi-temperature structure. The ratios measured in these two galaxies are significantly different from each other, which suggests that the resonant scattering effect in NGC 5813 is much stronger than in NGC 5044.

The comparison of the observed ratios with the line ratios from an optically-thin plasma calculated by the different spectral codes yields a very interesting result. In Fig.~\ref{fig:ratio_all}, the observed ratio for NGC 5044 appears to be below the predicted ratio from AtomDB v2.0.1 at a $\sim2\sigma$ confidence level considering only statistical errors. Since the $(I_{17.05} + I_{17.10})/I_{15.01}$ ratio can only increase due to resonant scattering in the core of an elliptical galaxy, we cannot explain the low value for NGC 5044 if the AtomDB v2.0.1 ratio is assumed. The value of 1.25 is, however, consistent with the predicted ratio using \citet{doron2002}. The elliptical galaxy \object{NGC 4649} in the sample of \citet{werner2009} and M87 \citep{werner2006b} also show a ratio of $\sim$1.25--1.30, but with a larger uncertainty. It should be noted that one of the points, \object{NGC 4636} out, was derived from a RGS spectrum just outside the galaxy centre where the optical depth of the 15~\AA~line is expected to be smaller than in the centre. 

Since the $(I_{17.05} + I_{17.10})/I_{15.01}$ ratio for an optically-thin plasma is uncertain from the atomic data side and the \citet{doron2002} results are problematic (Section~\ref{sec:atomic}), we are left without a good reference ratio to base our turbulence estimates on. The best we can do is to assume that the hot gas in \object{NGC 5044} is subject to such high levels of turbulence that the gas is effectively optically thin and estimate the turbulence in \object{NGC 5813} relative to that ratio. Since the shape of the $(I_{17.05} + I_{17.10})/I_{15.01}$ ratio as a function of temperature from \citet{doron2002} is very similar to the shape of the AtomDB v2.0.1 curve (see Fig.~\ref{fig:ratio_all}), and because it is also consistent with the ratio from \object{NGC 5044}, we assume for this paper that \citet{doron2002} indeed describe the \ion{Fe}{XVII} ratios in an optically-thin plasma correctly and that the gas in \object{NGC 5044} is optically thin. At the same time, we are aware of the problems with the \citet{doron2002} result. Therefore, we discuss the systematic uncertainties extensively.

\subsection{Turbulent velocity estimates}

The $(I_{17.05} + I_{17.10})/I_{15.01}$ line ratios from the fits can be directly compared to the predicted ratios from our Monte Carlo simulations. The observed ratios and the Monte Carlo results are plotted in Fig.~\ref{fig:resscat}. Note that the simulation results have been re-scaled to be relative to the optically-thin line ratio predicted using SPEX. The horizontal error bar on our data point indicates the RGS extraction region. From the simulations, it is evident that the predicted ratio varies substantially within this region. In order to make a fair comparison, we choose to calculate the emission-weighted average of the predicted line ratio within the $-0.5^{\prime}$--$0.5^{\prime}$ field of view. We use the count rates observed in the cross-dispersion direction of RGS in this interval as weighting factors for the calculation. The resulting ratios are listed in Table~\ref{tab:velocity}. 

\begin{table}[t]
\caption{Estimates for the turbulent velocities in NGC 5044 and NGC 5813.}
\label{tab:velocity}
\begin{tabular}{lrr|lrr}
\multicolumn{3}{c}{\textbf{NGC 5044}} 			& \multicolumn{3}{c}{\textbf{NGC 5813}} \\
\hline\hline
$M$		& $V_{\mathrm{turb}}$\tablefootmark{a}	
					& Pred.\tablefootmark{b}		
							&   $M$			& $V_{\mathrm{turb}}$\tablefootmark{a} 
												& Pred.\tablefootmark{b} \\
\hline
  0.00	        &   0   	        & 1.76	      	&   0.00  	      	&   0   	& 2.94  \\
  0.25       	& 115   	        & 1.53	      	&   0.25  	      	& 104 	   	& 2.13  \\
  0.50       	& 230   	        & 1.42	      	&   0.50  	      	& 208 	   	& 1.75  \\ 
  0.75       	& 345   	        & 1.37	      	&   0.75  	      	& 311 	   	& 1.61  \\
  $\infty$	& $\infty$		& 1.27		& $\infty$		& $\infty$	& 1.30  \\
\hline
\multicolumn{2}{l}{Observed}		& 1.25$\pm$0.08	&  \multicolumn{2}{l}{Observed}		& 1.87$\pm$0.08 \\
\hline 
\multicolumn{2}{l}{Lower Limit}		& $>$ 320 km s$^{-1}$	&  \multicolumn{2}{l}{Lower Limit}& $>$ 140 km s$^{-1}$ \\
\multicolumn{2}{l}{Upper Limit \tablefootmark{c}}	
					& $<$ 770 km s$^{-1}$	&  \multicolumn{2}{l}{Upper Limit \tablefootmark{c}}	
												& $<$ 540 km s$^{-1}$ \\
\hline
\end{tabular}
\tablefoot{
\tablefoottext{a}{Equivalent turbulent velocity estimate in km s$^{-1}$ using the Mach number and the adiabatic sound speed, which is 460 km s$^{-1}$ for NGC 5044 and 415 km s$^{-1}$ for NGC 5813.}
\tablefoottext{b}{Emission weighted average of the predicted $(I_{17.05} + I_{17.10})/I_{15.01}$ ratio over the RGS extraction region.}
\tablefoottext{c}{90\% upper limits on the turbulence derived from the line width in RGS.}
}
\end{table}

Using the emission weighted ratios from the simulation and the observed ratios in Table~\ref{tab:velocity}, we can constrain the level of the turbulent velocity in the inner 0.5$^{\prime}$ ($r \lesssim$ 5 kpc) of the galaxy. For NGC 5044, this is more difficult than for NGC 5813, because of our assumption that the gas in NGC 5044 is optically thin. At these low optical depths, the resonant scattering method loses sensitivity in the velocity estimate. Therefore we can only derive a lower limit for the turbulence. If we only consider statistical errors, the 90\% confidence lower limit is 320 km s$^{-1}$ for NGC 5044. For NGC 5813, the 15~\AA~line is significantly suppressed by resonant scattering with respect to the SPEX prediction, which allows a better estimate of the turbulence. To obtain the velocities in km s$^{-1}$, we interpolate linearly between the simulated values in Table~\ref{tab:velocity}. Since the errors due to the linear interpolation are smaller than the systematic errors in the atomic data, we do not gain accuracy by performing more Monte Carlo simulations for other Mach numbers. Our best estimate for $V_{\mathrm{turb}}$ in NGC 5813, without taking into account any systematic error, is 175$\pm$22 km s$^{-1}$, which means a statistical precision of about 13\%.   

From the discussion in Sect.~\ref{sec:resscat}, however, we know that we also need to take into account the systematic uncertainty in the optically-thin line ratios. For the calculation above, we used the line ratio predicted by SPEX, which appears to be a reasonable estimate. But, the typical difference between SPEX and AtomDB v2.0.1 is about 15\%, which also adds to the uncertainty. Using the AtomDB v2.0.1 ratio would not give a physical result for NGC 5044, but we can calculate it for NGC 5813. Assuming an optically-thin ratio of 1.513, the turbulent velocity would be 310 km s$^{-1}$. The difference between the velocities based on the AtomDB v2.0.1 and the SPEX ratio is a factor of two, which means that deriving the turbulent velocity in this way is very sensitive to the assumed optically-thin line ratio.   

Although the optically-thin line ratio is sensitive to the plasma temperature, we do not expect that multi-temperature structure in the gas adds to the systematic uncertainty. Fig.~\ref{fig:ratio_all} shows that the line ratio has its largest decline between 0.2 and 0.5 keV. From 0.5 to 1.0 keV, the ratio flattens. The temperatures that we measure for NGC 5044 and NGC 5813 are in the range 0.6--0.8 keV, where the ratio is fairly constant. The ratio could be biased if there were temperatures present below 0.5 keV. However, we do not detect significant \ion{O}{VII} emission in the spectrum, which would indicate the presence of those low temperatures. It is therefore safe to assume that the \ion{Fe}{XVII} line ratios are not biased due to a significant contribution from low temperatures and that the systematic uncertainty in the derived velocities due to multi-temperature structure is negligible.   

Another method of estimating the turbulence from high-resolution spectra is by measuring the line width. Although the line broadening due to the spatial extent of the source makes an accurate determination of the turbulent velocity impossible, it is possible to obtain an upper limit of the turbulent velocity with RGS. In the fit, we both allow the turbulent velocity in the CIE model and the scale factor $s$ that controls the width of the spatial distribution to be free. Since the spatial distribution is different from being Gaussian, it does not correlate significantly with the Gaussian velocity broadening, which allows us to obtain upper limits with a good accuracy. During the error calculation, also the scale factor $s$ is allowed to optimise, which means that uncertainty is included in the upper limit. The results are shown on the last row of Table~\ref{tab:velocity}. For NGC 5044, the 90\% upper limit constrains the turbulent velocity to the range between 320 $< V_{\mathrm{turb}} <$ 770 km s$^{-1}$. The upper limit for NGC 5813 of $<$ 540 km s$^{-1}$ is consistent with the estimated value from resonant scattering.  

An important question to understand the structure of the ICM in groups and clusters of galaxies is what fraction of the pressure in the ICM is due to turbulence. This fraction is equivalent to the ratio of the turbulent energy ($\epsilon_{\mathrm{turb}}$) to the thermal energy ($\epsilon_{\mathrm{therm}}$), which can be easily calculated from the Mach number of the turbulence \citep{werner2009}: 
\begin{equation}
\frac{\epsilon_{\mathrm{turb}}}{\epsilon_{\mathrm{therm}}} = \frac{\gamma}{2} M^2, 
\end{equation}
where $\gamma$ is the adiabatic index for an ideal gas ($\gamma = 5/3$). From the lower limit derived for NGC 5044, we derive that the turbulent energy can range from 40\% of the thermal energy up to a factor of 2 times the thermal energy. The turbulent pressure clearly is a component which should be considered for NGC 5044. The turbulent energy fraction for NGC 5813 is significantly lower, between 15\% and 45\% if we take into account the systematic uncertainty in the initial line ratio. Since we only probe the inner 0.5$^{\prime}$ of both galaxies with RGS and the optical depth in the 15 \AA~line decreases rapidly above $r >$ 10 kpc, the values derived above are valid for the inner regions within $r \lesssim$ 5--10 kpc.     

\subsection{Chandra images of NGC 5044 and NGC 5813}

\begin{figure*}[t]
\begin{center}
\includegraphics[width=1.01\columnwidth]{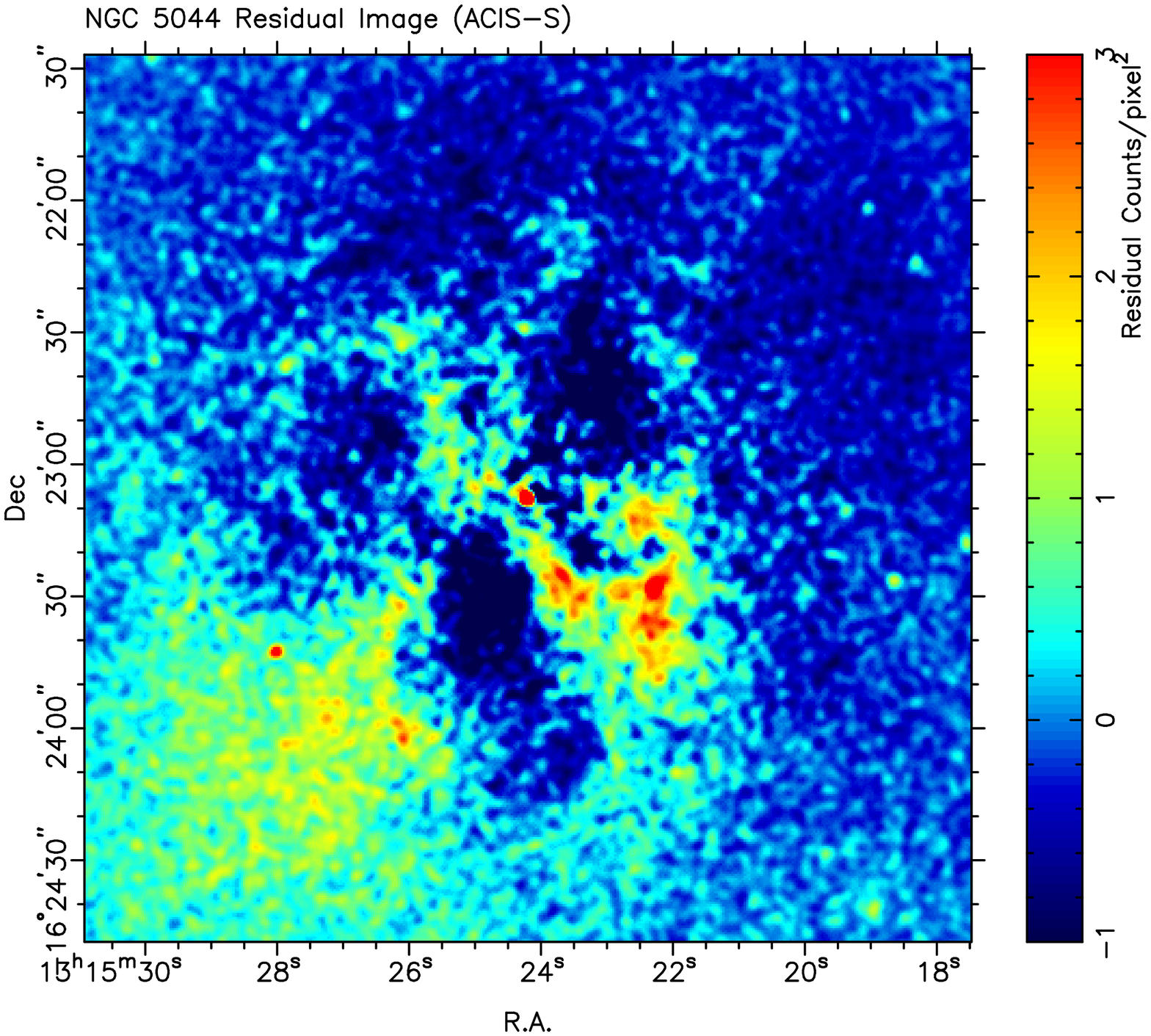}
\includegraphics[width=1.01\columnwidth]{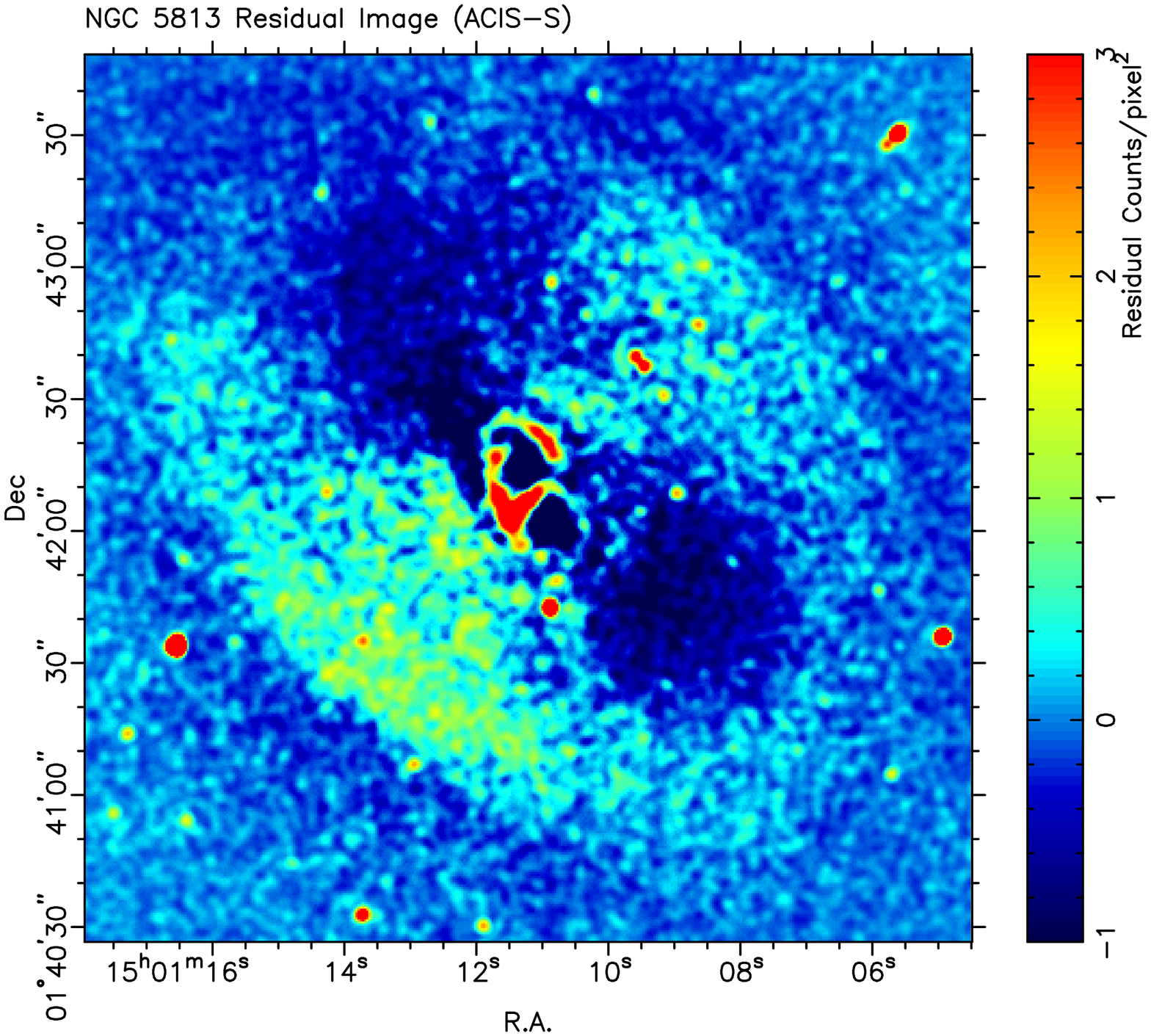}
\end{center}
\caption{{\it Chandra} ACIS images of NGC 5044 ({\it left}) and NGC 5813 ({\it right}) after subtraction of the average ICM emission.}
\label{fig:residual}
\end{figure*}

Since we find a significant difference between the measured turbulent velocities in NGC 5044 and NGC 5813, it is interesting to look for the origins of the turbulence by looking at the spatial structure of the groups. They both show X-ray cavities blown by the central AGN in their core, which is a possible source of turbulence. In order to visualise these cavities, we used the 4 {\it Chandra} ACIS observations listed in Table~\ref{tab:chandra} to create an image where the average ICM emission is subtracted \citep[see also e.g.][for NGC 5044 and NGC 5813 respectively]{david2009,randall2011}. In all observations, the source was centred on ACIS-S3 of the ACIS-S detector. For both sources, the two individual event files from the standard pipeline products were merged to create a single event file containing the maximum possible exposure. Then, we extracted radial surface brightness profiles within a 2.5$^{\prime}$ radius from the central AGN in the 0.5 -- 7 keV band. The maximum extraction radius of 2.5$^{\prime}$ falls within the CCD 7 boundaries in all observations. The derived surface brightness profiles were fitted using an empirical model consisting of two King profiles and a constant. The full expression for this model $f(r)$ is:
\begin{equation}
f(r) = \frac{N_1}{\left[1 + \displaystyle \frac{r^2}{r_{c1}^2}\right]^{\beta_1}} + \frac{N_2}{ \left[1 + \displaystyle \frac{r^2}{r_{c2}^2}\right]^{\beta_2}} + C
\label{eq:dblking}
\end{equation}
with normalisations $N_1$ and $N_2$, core radii $r_{c1}$ and $r_{c2}$, and exponents $\beta_1$ and $\beta_2$ for both King profiles. A constant $C$ was added to account for the background emission. The expected number of counts from the best-fit double-King model was subsequently subtracted from the observed number of counts in the original image for every pixel. The residual images that we obtained were smoothed with a Gaussian using a width ($\sigma$) of 0.98$^{\prime\prime}$. The result for both galaxies is shown in Fig.~\ref{fig:residual}.

\section{Discussion}

Using deep {\it XMM-Newton} RGS data of \object{NGC 5044} and \object{NGC 5813}, we have found a significant difference between the $(I_{17.05} + I_{17.10})/I_{15.01}$ line ratios in these galaxy cores. Our result shows that the \ion{Fe}{XVII} line at 15 \AA~in NGC 5813 is significantly affected by resonant scattering, confirming the 2$\sigma$ hint found by \citet{werner2009} that is based on a shorter 30 ks RGS exposure. The $(I_{17.05} + I_{17.10})/I_{15.01}$ line ratio in NGC 5044, on the other hand, is found to be consistent with the line ratio predicted by the CIE model in SPEX, which is in agreement with the conclusion of \citet{tamura2003}, based on a 20 ks RGS exposure, that resonance scattering effects in NGC 5044 are small. The difference between the line ratios in these galaxies is best explained by resonant scattering in the 15 \AA~line of NGC 5813, which is strong due to the relatively modest turbulent velocities ($\sim$0.4$M$) in the core of this object. Other effects that could affect the measured line ratio, like multi-temperature structure, do not appear to play a role. Interestingly, we do not find strong evidence for multi-temperature structure in NGC 5044, contrary to the results found by \citet{tamura2003} and \citet{grange2011}. This difference originates from our relatively narrow cross-dispersion selection that we need to detect resonant scattering. We extract the RGS spectrum from a 1$^{\prime}$ wide region in the cross-dispersion direction, while \citet{tamura2003} use a width of 2$^{\prime}$ and \citet{grange2011} a width of 5$^{\prime}$. The origin of the multi-temperature signal reported in these previous works is therefore most likely the temperature gradient in the inner few arcminutes of the source.

\subsection{Systematics in the optically-thin line ratio}

The systematic uncertainty in the unscattered optically-thin $(I_{17.05} + I_{17.10})/I_{15.01}$ ratio is the most important limitation for measuring turbulent velocities in clusters and elliptical galaxies. For NGC 5813, we find a difference of a factor of two in the derived turbulent velocity when the optically-thin line ratio is increased by only 15\%. This clearly adds a large systematic uncertainty to the derived turbulence. The result can only be improved when the uncertainty in the optically-thin line ratio decreases, which is difficult due to the problems with the atomic data of \ion{Fe}{XVII}. The estimated uncertainty on individual line strengths produced by spectral codes is about 20--30\%. 

RGS observations of elliptical galaxies like NGC 5044 also help to constrain the optically-thin line ratio. Although we currently lack the statistics to definitively rule out certain model calculations, a larger sample of deep observations of these objects will help to constrain the upper limits for this ratio. As shown in Section~\ref{sec:atomic}, laboratory measurements and theoretical calculations could eventually improve our knowledge about the optically-thin line ratio as soon as they yield consistent results. There are signs that theoretical and laboratory results are converging \citep{gillaspy2011}, but this still needs confirmation. Considering the remaining uncertainties in individual line strengths, it is clearly necessary to invest in theoretical and experimental research to improve the atomic data before new missions like {\it Astro-H} are launched \citep{takahashi2010}. This work shows that accurate atomic data is important to perform robust measurements of astrophysical sources.

\subsection{Line profiles and velocity broadening}

Upper limits on the turbulent velocity were estimated for NGC 5044 and NGC 5813 before by \citet{sanders2011} based on the oldest data sets with unfiltered exposures of 43 and 73 ks, respectively. They find upper limits of 1500 km s$^{-1}$ for NGC 5044 and 1400 km s$^{-1}$ for NGC 5813 at the 90\% confidence level. Our upper limits are significantly smaller. For NGC 5044, we find an upper limit based on line broadening of $<$770 km s$^{-1}$, which is a factor of $\sim$1.95 improvement with respect to \citet{sanders2011}. This is consistent with the expected improvement ($\sim\sqrt{3}$) due to the increased statistics. For NGC 5813, the improvement is much larger than expected based on statistics. The upper limit drops from $<$1400 km s$^{-1}$ to $<$540 km s$^{-1}$ (factor $\sim$2.5), which is much better than the improvement of the statistics (factor $\sqrt{2}$). We note that the method of correcting for the spatial broadening is different in \citet{sanders2011}, therefore the results may not be fully comparable. Although our upper limit for NGC 5813 is lower than expected, it is still consistent with the independent measurement using resonant scattering.     

We must also realise that the velocity broadening measurements of elliptical galaxies with RGS are sensitive to systematic effects. The line profile of a line in RGS does not only depend on the velocity broadening, but also on the line strength as a function of radius in the elliptical galaxy. Since the line strength of a certain line depends on the temperature, density, element abundance, and resonant scattering properties, the line profile observed with RGS can be different for each line. In order to include the bias due to the different line widths in the upper limit, we allow the width of the broadening profile derived from MOS ($s$) to optimise. The derived upper limits on the turbulence thus give a rough estimate of what level of turbulence would still be consistent within the average of the line widths of the lines in the spectrum.       

In addition, resonant scattering would also change the profile of the 15 \AA~line. The core of the line would be suppressed and because resonant scattering is a conservative process, the line wings should be enhanced. Unfortunately, line blends prevent us from studying the wings of the 15 \AA~line in more detail. Since the strongest resonant scattering effect is the suppression of the core of the line, our method of using delta lines convolved with the spatial line profile give us a good estimate of the line ratios. The other ICM properties that determine the line profile, like temperature, density, and abundance do not influence our estimate of the $(I_{17.05} + I_{17.10})/I_{15.01}$ ratio, because these lines are all emitted by \ion{Fe}{XVII} and they should have a similar profile in the observed temperature range, apart from resonant scattering effects.   

\subsection{Origin of turbulence}

The RGS spectra of NGC 5044 and NGC 5813 were extracted from the bright inner regions ($r \lesssim$ 5 kpc) of both galaxies. The optical depth in the 15 \AA~line declines rapidly beyond a radius of $\sim$10 kpc. It is therefore reasonable to assume that most of the resonant scattering signal originates from the inner $\lesssim$10 kpc of the galaxy. Strictly speaking, any differential gas motion can contribute to the line broadening and affect the optical depth. Therefore, our estimate of the turbulence and the turbulent energy implicitly includes all types of different motions, provided that they happen on spatial scales where the accumulated optical depth is $\gtrsim$1. If one assumes normal turbulence (e.g. with a Kolmogorov spectrum) most of the energy is found at the largest scales. Only eddies with a characteristic size comparable to or smaller than the size of the optically thick core affect the amplitude of the resonant scattering significantly \citep{zhuravleva2011}.

Both \object{NGC 5044} and \object{NGC 5813} show AGN activity in the form of bubbles, therefore the considerable difference in the measured turbulence may be difficult to explain by AGN induced turbulence alone. The AGN activity in \object{NGC 5813} has produced bubbles and relatively strong shocks \citep{randall2011}, but apparently only modest differential gas motions and turbulence along the line of sight compared to NGC 5044. The inner 10 kpc of NGC 5044, on the other hand also shows evidence of recent $\sim10^7$ yr AGN activity \citep{gastaldello2009}, but no evidence of strong shocks like in \object{NGC 5813}. Therefore, the AGN activity in NGC 5813 appears to be stronger than in NGC 5044, which is opposite to the conclusion that one would reach based on the limits on the turbulence that we obtained. AGN activity alone cannot explain the difference in turbulence that we detect between \object{NGC 5044} and \object{NGC 5813}.

Another source of turbulence would be cold fronts induced by an off axis merger. {\it Chandra}, {\it XMM-Newton}, and optical data show evidence that NGC 5044 had a recent merger with a small sub group \citep{mendel2008,gastaldello2009,david2009}. The {\it Chandra} images of NGC 5044 and NGC 5813 in Fig.~\ref{fig:residual} indeed show a striking difference. While the bubbles in NGC 5813 appear to rise in a straight line along the NE--SW axis, the bubbles in NGC 5044 rise in many directions. Most likely, the recent merger in NGC 5044 has induced a sloshing, or swirling, motion in its ICM, causing the bubbles to rise in a rotating medium \citep{david2009}. Along the edges of the cold fronts associated with the merger, instabilities cause turbulent motions that may be detected using resonant scattering. A potential problem with this interpretation is the fact that numerical simulations, like e.g. \citet{ascasibar2006} and \citet{roediger2011} show that for a sloshing cluster the velocity field in the centre of the hot gas has a much smaller amplitude than the velocities in the outskirts. Since we are probing mainly the inner $\sim$10 kpc of the galaxy, we should not see the high velocities associated with the merger. On the other hand, these simulations were made for larger systems and different merger geometries. We cannot exclude the possibility that in NGC 5044 the gas in the centre is disturbed by the sloshing motions. NGC 5813 does not appear to be disrupted by a recent merger \citep{randall2011}. Most likely differential gas motions, including turbulence, due to the gas sloshing are responsible for the observed small optical depth of the \ion{Fe}{XVII} line at 15 \AA.  

The derived values for the turbulent energy fraction in NGC 5813 (15--45\%), assuming isotropic turbulence, are comparable to estimates from numerical simulations \citep[e.g.][]{dolag2005,nagai2007}, which range between 5--30\% of the thermal energy. The turbulent velocities around 100--200 km s$^{-1}$ could very well be induced by the bubbles blown by the AGN. \citet{heinz2010} show that bubbles can generate turbulence as high as $\sim$500 km s$^{-1}$ in clusters of galaxies. Due to the reduced scale of an elliptical galaxy with respect to a cluster, an AGN generated velocity of 100--200 km s$^{-1}$ seems reasonable. NGC 5044, on the other hand, shows a very high turbulent energy fraction $>$40\%. The ICM of NGC 5044 appears to be very disturbed. Due to the low mass of NGC 5044, a merger is the most likely explanation for the high level of turbulence.

\subsection{Future prospects}

Estimating turbulent velocities through resonant scattering and line broadening, like presented in this paper, is at the limit of the current instrument capabilities. Although fitting high-resolution spectra from new micro-calorimeter instruments, like SXS aboard {\it Astro-H} \citep{takahashi2010}, will be a challenge for the current atomic databases and spectral codes, the excellent spectral resolution of SXS will enable us to measure turbulent velocities directly from the line widths. Since the spectral band of SXS ranges from 0.3--12 keV, it will also allow the study of resonant scattering in other strong resonance lines, for example in the Fe-K region around 6 keV \citep[see the review by][]{churazov2010}. The combined effort of improving atomic data and flying new instruments like SXS aboard {\it Astro-H} will enable a more comprehensive study of turbulence in elliptical galaxies and clusters of galaxies. The high spectral resolution will not only allow  measurements of the amplitude of the motions, but also their anisotropy and spatial scales.

\section{Conclusions}

We analyse deep {\it XMM-Newton} RGS spectra of \object{NGC 5044} and \object{NGC 5813}. Following a detailed analysis of the $(I_{17.05} + I_{17.10})/I_{15.01}$ line ratio, we find that:
\begin{itemize}
\item The $(I_{17.05} + I_{17.10})/I_{15.01}$ ratio in NGC 5813 (1.87$\pm$0.08) is significantly higher than in NGC 5044 (1.25$\pm$0.08). We interpret this difference to be line suppression of the \ion{Fe}{XVII} line at 15 \AA~due to resonant scattering, which strongly depends on the turbulent velocities in the ICM.
\item The core of NGC 5044 shows a $(I_{17.05} + I_{17.10})/I_{15.01}$ line ratio that is consistent with the theoretical ratio by \citet{doron2002}. It is, however, 2$\sigma$ below the AtomDB v2.0.1 ratio, which cannot be explained by resonant scattering. Although the distorted wave calculations by \citet{doron2002} are not state-of-the-art, it is consistent with our ratio in NGC 5044. Therefore, we assume this optically-thin line ratio to estimate the turbulent velocities in NGC 5044 and NGC 5813. 
\item Using the measured $(I_{17.05} + I_{17.10})/I_{15.01}$ ratio, we estimate the level of turbulence within $r \lesssim$ 5--10 kpc of the centre of these galaxies assuming isotropic turbulence. The core of NGC 5813 shows moderate turbulent velocities between 140$<V_{\mathrm{turb}}<$540 km s$^{-1}$ (90\% confidence) with a best estimate of $\sim$175 km s$^{-1}$. The ratio of the turbulent energy over the thermal energy, which is the same as the fraction of the turbulent pressure support, yields a turbulent energy fraction between $\sim$15--45\%, which is comparable to estimates of the turbulent pressure support from numerical simulations. The lower and upper limit for the turbulence in NGC 5044 yields a velocity range of 320$<V_{\mathrm{turb}}<$770 km s$^{-1}$, which means a turbulent energy fraction of $>$ 40\%.

\end{itemize}

\begin{acknowledgements}
This work is based on observations obtained with XMM-Newton, an ESA science mission
with instruments and contributions directly funded by ESA Member States and
NASA. The Netherlands Institute for Space Research (SRON) is supported financially
by NWO, the Netherlands Organisation for Scientific Research. 
I. Zhuravleva would like to thank the International Max Planck Research School 
on Astrophysics (IMPRS) in Garching.
N. Werner is supported by the National Aeronautics and Space Administration through 
Chandra/Einstein Postdoctoral Fellowship Award Numbers PF8-90056 and PF9-00070 issued 
by the Chandra X-ray Observatory Center, which is operated by the Smithsonian Astrophysical 
Observatory for and on behalf of the National Aeronautics and Space Administration under contract NAS8-03060.
\end{acknowledgements}

\bibliographystyle{aa}
\bibliography{clusters}

\end{document}